\documentclass[aps,10pt,prl,twocolumn,showpacs,groupedaddress]{revtex4}  
\usepackage{graphicx}  
\usepackage{dcolumn}   
\usepackage{bm}        
\usepackage{amssymb}   
\usepackage{amsmath}

\begin{document}

\title{Anomalous second order coherence and $g^{(2)}$ complementarity}
\author{Ron Folman}
	\email{folman@bgu.ac.il}

	\affiliation{Department of Physics, Ben-Gurion University of the Negev, Be'er Sheva 84105, Israel}
\date{\today}

\begin{abstract}
This paper is a summary of my talk at SPIE2013. The organizers were kind enough to invite me to talk about anything I wanted, and I chose to bring up the notion of higher order complementarity and the fact that it may not be monotonic. I enter this discussion, which is rather speculative at this stage, by calculating a specific example. This is to be regarded as work in progress.

We analyze a two-particle state and show that when mismatching the detector frequency response and the field frequency spectrum, several anomalous features become apparent. In particular, while we find several well known features, such as $g^{(2)}(\tau=0)=1/2$ for completely indistinguishable particles, we also find that as the photons are slightly separated and may be distinguished, $g^{(2)}$ may oscillate as a function of the detector bandwidth. Beyond the latter interesting observation for which we cannot find a simple physical origin, we also suggest, an interpretation with regard to an oscillating $g^{(2)}$ complementarity.
\end{abstract}

\maketitle

Intensity correlations known as the second-order coherence function $g^{(2)}$, stand as one of the most measured observables in quantum optics, and have given rise to numerous fundamental insights.
The $g^{(2)}$ observable received its fame mostly in the spatial domain, through work on the Hanbury Brown and Twiss effect \cite{HBT1954,Fano1961,Baym1969,HBT1974} and the 2-photon Hong-Ou-Mandel effect \cite{Hong1987,Ou1989,Pittman1996,Mandel,Boyd2008} (for an overview see the book of Scully and Zubairy \cite{Scully}; see also our work on an anomalous transmission for 2-photon states through a 3-port junction \cite{ron}). In the time domain, on which we focus here, it seems the 2-photon state has received less attention, though considerable work has been done \cite{shih,Rempe,Ou2006}. A typical text book plot of the  $g^{(2)}(\tau)$ function exhibits three lines representing a thermal or chaotic source ($g^{(2)}(0)=2$), a coherent or laser source ($g^{(2)}(0)=1$) and a single particle source, e.g. a single atom ($g^{(2)}(0)=0$). All three lines converge to $g^{(2)}(\tau\gg0)=1$. Recently, renewed experimental and theoretical attention has been given to the interesting statistical properties of the $g^{(2)}(\tau)$ function for a two-particle state, e.g. recent work by Rosencher, Fabre and colleagues \cite{Emmanuel2009,Emmanuel2012}, and by Elena del Valle and colleagues \cite{Elena1,Elena2,Elena3}. Here we present several anomalous features of second-order coherence. For example, we show for the first time that it may oscillate as a function of the detector bandwidth. Furthermore, we suggest that one may connect this rather strange result to the realm of complementarity.

Our observable hence forth will be the $g^{(2)}(\tau)$ temporal second-order coherence function, which may be measured by a single detector observing a channel of incoming particles. Let us describe the experimental procedure. A source gives rise to a photon pair where the two photons are labeled $i$ and $j$. The source may be made of a down converter (e.g. \cite{scully}) or of two adjacent atoms which are excited by a brief laser pulse (e.g. \cite{grainger}). Different sources obviously give rise to different levels of {\it indistinguishability}, a feature which we will later quantify by a parameter $J$.  We denote the incoming state as $|11\rangle$ rather than $|2\rangle$ to emphasize the possible distinguishability.

Before embarking on our specific calculation, let us remind ourselves that for two identical particles $g^{(2)}(\tau=0)=1/2$. This is indeed simple to calculate. Using $|11\rangle=\frac{1}{\sqrt{2!}}(\hat{A}^\dag)^2|0\rangle$ ($|0\rangle$ is the vacuum state) and the commutation relation $[\hat{A},\hat{A}^\dag]=1$, and noting that $\langle 11| \hat{A}^\dag\hat{A}|11\rangle=2$, the value of this correlation function is:
\begin{eqnarray}
\frac{\langle 11| \hat{A}^\dag\hat{A}^\dag\hat{A}\hat{A}|11\rangle }{\langle 11| \hat{A}^\dag\hat{A}|11\rangle ^2}=
\frac{1/2\langle0|(\hat{A})^2 \hat{A}^\dag\hat{A}^\dag\hat{A}\hat{A}(\hat{A}^\dag)^2|0\rangle }{2 ^2}=1/2.
\label{eq:g2b}
\end{eqnarray}

It is worthwhile noting that introducing a $50/50$ beam-splitter (BS) and a second detector, as is done in many experiments due to detector limitation, gives the same result. In the following we will therefore use a the single mode calculation ($\hat{A}^\dag_1$).

Let us now introduce realistic photons by specifying a frequency spectrum for the incoming state, where we shall utilize Loudon's formalism \cite{BS}. We define the creation operators of the two photons as
\begin{eqnarray}
\hat{A}_{1i}^\dag=\int d\omega \alpha_i^*(\omega)\hat{a}_1^\dag(\omega)~~~~
\hat{A}_{1j}^\dag=\int d\omega \alpha_j^*(\omega)\hat{a}_1^\dag(\omega),
\label{eq:incoming1}
\end{eqnarray}
where $\alpha_i(\omega)$ and $\alpha_j(\omega)$ are the normalized spectral amplitudes for the two photons. The overlap integral of these functions is
\begin{eqnarray}
J=\int d\omega \alpha_i(\omega)\alpha_j^*(\omega).
\label{eq:incoming2}
\end{eqnarray}

The operator $\hat{a}_1^\dag(\omega)$ creates a photon and satisfies the commutation relation $[\hat{a}_1(\omega),\hat{a}_1^\dag(\omega')]=\delta(\omega-\omega')$. The field operators defined in Eq. (\ref{eq:incoming1}) have commutation relations $[\hat{A}_{1i},\hat{A}_{1i}^\dag]=[\hat{A}_{1j},\hat{A}_{1j}^\dag]=1$ and $[\hat{A}_{1i},\hat{A}_{1j}^\dag]=[\hat{A}_{1j},\hat{A}_{1i}^\dag]^*=J$, with the origin of the latter being
\begin{eqnarray}
[\hat{A}_{1i},\hat{A}_{1j}^\dag]=
\int d\omega \int d\omega' |\alpha_i(\omega)||\alpha_j^*(\omega')|[\hat{a}_1(\omega),\hat{a}^\dag_1(\omega')]\nonumber\\
e^{-i(\omega_0-\omega)t_i} e^{i(\omega_0-\omega')t_j}=\nonumber\\
\int d\omega \int d\omega' |\alpha_i(\omega)||\alpha_j^*(\omega')|\delta(\omega-\omega')\nonumber\\
e^{-i(\omega_0-\omega)t_i} e^{i(\omega_0-\omega')t_j}=\nonumber\\
\int d\omega |\alpha(\omega)|^2 e^{i(\omega_0-\omega)\tau}=J(\tau),
\label{eq:incoming7b}
\end{eqnarray}
where $\tau=t_j-t_i$ and where we have taken the spectral amplitude of both photons to be
\begin{eqnarray}
\alpha_{i,j}(\omega)=
|\alpha(\omega)|\text{exp}[-i(\omega_0-\omega)t_{i,j}],
\label{eq:alpha}
\end{eqnarray}
where $\omega_0$ is the photon center frequency. A typical photon spectral amplitude is (e.g. \cite{Loudon})
\begin{eqnarray}
\alpha_{i,j}(\omega)=(2\pi\Delta^2)^{-\frac{1}{4}}\text{exp}[-i(\omega_0-\omega)t_{i,j}
-\frac{(\omega_0-\omega)^2}{4\Delta^2}]=\nonumber\\
|\alpha(\omega)|\text{exp}[-i(\omega_0-\omega)t_{i,j}],
\label{eq:alpha2}
\end{eqnarray}
where $\Delta$ is the frequency spectrum width.

As $\langle0|\hat{A}_{1j}\hat{A}_{1i}\hat{A}_{1i}^\dag\hat{A}_{1j}^\dag|0\rangle=1+|J|^2$ we may write the normalized input state as
\begin{eqnarray}
|11\rangle =(1+|J|^2)^{-1/2}\hat{A}_{1i}^\dag\hat{A}_{1j}^\dag|0\rangle.
\label{eq:incoming3}
\end{eqnarray}

If we introduce varying levels of distinguishability between the photons,
we find for the numerator in Eq. (\ref{eq:g2b}):
\begin{eqnarray}
\langle11| \hat{A}_{1j}^\dag\hat{A}_{1i}^\dag\hat{A}_{1i}\hat{A}_{1j}|11\rangle=\nonumber\\
(1+|J|^2)^{-1}\langle0|\hat{A}_{1j}\hat{A}_{1i} \hat{A}_{1j}^\dag\hat{A}_{1i}^\dag\hat{A}_{1i}\hat{A}_{1j}\hat{A}_{1i}^\dag\hat{A}_{1j}^\dag|0\rangle=\nonumber\\
\frac{1+2|J|^2+|J|^4}{(1+|J|^2)}=(1+|J|^2).
\label{eq:g2d}
\end{eqnarray}

Similarly, the denominator is:
\begin{eqnarray}
\langle11|\hat{A}_{1i}^\dag\hat{A}_{1i}|11\rangle \langle11|\hat{A}_{1j}^\dag\hat{A}_{1j}|11\rangle=\nonumber\\
(1+|J|^2)^{-2}\langle0|\hat{A}_{1j}\hat{A}_{1i}\hat{A}_{1i}^\dag\hat{A}_{1i}\hat{A}_{1i}^\dag\hat{A}_{1j}^\dag|0\rangle\times\nonumber\\
\langle0|\hat{A}_{1j}\hat{A}_{1i}\hat{A}_{1j}^\dag\hat{A}_{1j}\hat{A}_{1i}^\dag\hat{A}_{1j}^\dag|0\rangle=\nonumber\\
(1+|J|^2)^{-2}(1+3|J|^2)^2.
\label{eq:dis2}
\end{eqnarray}

\begin{figure}[b]%
\includegraphics[width=\columnwidth]{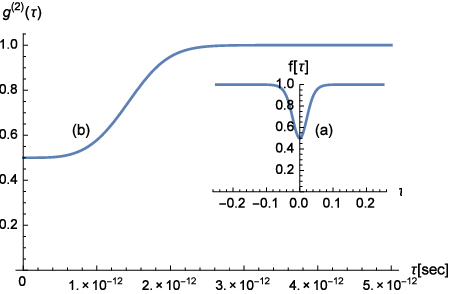}%
\caption{
$g^{(2)}(\tau)$. (a) Eq. (\ref{eq:dis3}) with $J(\tau)=e^{-(2\pi)^2|\tau|\Delta}$ (i.e. Lorentzian spectral distribution), where $\Delta$, the width of the spectral distribution, is taken to be one. As expected, the function has $g^{(2)}(0)=1/2$ and converges to $g^{(2)}(\tau)=1$ for larger times. We show both sides of the time axis to show that $g^{(2)}(\tau)$ is a symmetric function. This is not always necessarily so. (b) Eq. (\ref{final})  with $\Delta=\Gamma= 10^{12}$. We see that when $\tau > \frac{1}{\Delta}\approx\frac{1}{\Gamma}$ we have $g^{(2)}(\tau)=1$. This is as expected from Eq. (\ref{final}) as only $J_2$ and $J_5$ are non-zero in this regime. In the other limit, all $J$ parameters are one and as expected from Eq. (\ref{final}), $g^{(2)}(\tau)=1/2$.
}%
\label{PhotonCounting}%
\end{figure}

Hence, we now find the second order correlation to be
\begin{eqnarray}
g^{(2)}=(1+|J|^2) / (1+|J|^2)^{-2}(1+3|J|^2)^2 =\nonumber\\
(1+|J|^2)^3 / (1+3|J|^2)^2 ,
\label{eq:dis3}
\end{eqnarray}
giving $g^{(2)}(J=0)=1$, $g^{(2)}(J=1/2)\approx 0.6$ and $g^{(2)}(J=1)=1/2$, and zero derivatives at $J=0$ and $J=1$. In Fig. \ref{PhotonCounting}(a) we plot $g^{(2)}(\tau)$ based on $J(\tau)$ from Eq. (\ref{eq:incoming7b}), and we find that our model retrieves the expected form for $g^{(2)}(\tau)$.

In the above calculations the same operators are assumed for the field and the detection, while the frequency spectrum of the former and the frequency response function (bandwidth) of the latter, may be quite different. Furthermore, the time separation between the prepared photons and between the detection events may also be different. Let us therefore define the annihilation operator for the corresponding two detections $m=1,2$ as
\begin{eqnarray}
{\hat D}_{1m}=\int d\omega {\hat a}_1(\omega)\alpha^d_m(\omega),
\label{eq:new3}
\end{eqnarray}
where $\alpha^d_m(\omega)$ is the spectral response of the detector:
\begin{eqnarray}
\alpha^d_m(\omega)=(2\pi\Gamma^2)^{-\frac{1}{4}}\text{exp}[-i(\omega^d_0-\omega)t_m
-\frac{(\omega^d_0-\omega)^2}{4\Gamma^2}]=\nonumber\\
|\alpha^d_m(\omega)|\text{exp}[-i(\omega^d_0-\omega)t_m],
\label{eq:alphad}
\end{eqnarray}
where $\omega^d_0$ is the center frequency of the detector response function and $\Gamma$ is its bandwidth.
We can safely assume that $|\alpha^d_m(\omega)|$ is the same for any detection $m$. If we further assume that there is no unknown temporal smearing in the photoelectron ionization process, $1/\Gamma$ is the temporal resolution of the detector, while $\omega^d_0$ is the frequency in which the probability of photo-ionization is highest (i.e. the center frequency of the detector).

In simple terms, we are gating our detector so that we allow two detection windows with width $1/\Gamma$ and with a time separation of $t_1-t_2$. Once we fix the initial time difference between the photons and their spectral width, and we also determine the above gating of our detectors (time and spectral response), we have fixed our preparation and may then ask what is the $g^{(2)}$ value we should expect.

In the following we will use a frequency mismatch between the detector center frequency and that of the photon to create a de-facto vacuum, in which extremely small detection probabilities create a negligible denominator in Eq. (\ref{eq:new7}), enabling very large $g^{(2)}$ values. As the photons are created in pairs, the created state may be viewed as a bunched or squeezed - vacuum, and indeed it is well known that squeezed vacuum may give rise to very high $g^{(2)}$ values \cite{Loudon}.

Before continuing, some cautionary statements are warranted. Under the extreme conditions described in this paper, it may be that our simplified model for the detector (and perhaps even the source) should in future work be expanded with more detailed models in which the detection dynamics play a role. These extreme conditions include among other things, a ``sensitivity chirp" in which the large mismatch noted above, causes the photon to be detected on the far tail of the detector response function, and consequently there is a much higher sensitivity to one side of the photon frequency distribution relative to the other. Having said that, let us continue with our model and analyze the results.


For the single mode configuration (one detector and no BS) we can now write,
\begin{eqnarray}
g^{(2)}(\tau)=
\frac{\langle \hat{D}_{12}^\dag\hat{D}_{11}^\dag\hat{D}_{11}\hat{D}_{12}\rangle }{\langle \hat{D}_{11}^\dag\hat{D}_{11}\rangle \langle \hat{D}^\dag_{12}\hat{D}_{12}\rangle},
\label{eq:new7}
\end{eqnarray}
where the brackets indicate the expectation value for the input state as defined in Eq. (\ref{eq:incoming3}). As this is exactly what is written in Eqs. (\ref{eq:g2d})-(\ref{eq:dis2}), the result of Eq. (\ref{eq:dis3}) still holds.

Let us now differentiate between the frequency spectrum of the detector and that of the incoming photons (both in center frequency and width), and define six overlap parameters $J$ that describe the commutation relations between the photon operators and themselves, between the photon and detector operators, and between the detector operators and themselves such that $[\hat{A}_{1i},\hat{A}_{1j}^\dag]=J_1$, $[\hat{A}_{1i},\hat{D}_{11}^\dag]=J_2$, $[\hat{A}_{1i},\hat{D}_{12}^\dag]=J_3$, $[\hat{A}_{1j},\hat{D}_{11}^\dag]=J_4$, $[\hat{A}_{1j},\hat{D}_{12}^\dag]=J_5$ and $[\hat{D}_{11},\hat{D}_{12}^\dag]=J_6$.
Using these commutation relations we can now evaluate Eq. (\ref{eq:new7}) to give
\begin{eqnarray}
g^{(2)}(J_i)=(1+|J_1|^2)\times\nonumber\\
\frac{|J_2|^2|J_5|^2+|J_3|^2|J_4|^2+2Re\left[J_2^*J_3J_4J_5^*\right]}
{(|J_2|^2+|J_4|^2+2Re\left[J_1^*J_2J_4^*\right])(|J_3|^2+|J_5|^2+2Re\left[J_1^*J_3J_5^*\right])},
\label{final}
\end{eqnarray}
where we have verified that the same equation holds for a balanced configuration (two detectors and a BS).

Let us now analyze this result.
To begin with, we assume as before that the frequency spectrum of the two photons is identical, and that the frequency response in the two detection events is also the same. For simplicity, we first assume that the spectral widths of the photon wavepackets $\Delta$ and the detector response function $\Gamma$ are equal and constant, and that the center frequency of both entities is equal i.e. $\omega^d_0=\omega_0$. More over, we assume that the time difference $\tau$ is always the same for the incoming photons and the times of the measurements. Under these conditions one finds that $J_2=J_5=1$ and $J_1J_3=J_1J_4=|J_1|^2$, and consequently, Eq. (\ref{final}) [drawn in Fig. \ref{PhotonCounting}(b)] reduces to Eq. (\ref{eq:dis3}), as it should.

Let us generalize the above result further by fixing the time difference between the photons (i.e. fixed source parameters) to be $\tau^p=t_j-t_i$, while $\tau=t_2-t_1$ is the time between detections. Without loss of generality, let us define that the time of the first photon and the first detection are the same, namely, $t_1=t_i=0$ (e.g. $\tau^p-\tau=t_j-t_2$). We now find \cite{generality}
\begin{eqnarray}
J_1(\tau^p)=\int d\omega |\alpha(\omega)|^2e^{[-i(\omega-\omega_0)\tau^p]},\nonumber\\
J_2=\int d\omega |\alpha_i(\omega)||\alpha^{d*}_1(\omega)|,\nonumber\\
J_3(\tau)=\int d\omega |\alpha_i(\omega)||\alpha^{d*}_2(\omega)|e^{[-i(\omega-\omega^d_0)\tau]},\nonumber\\
J_4(\tau^p)=\int d\omega |\alpha_j(\omega)||\alpha^{d*}_1(\omega)|e^{[i(\omega-\omega_0)\tau^p]},\nonumber\\
J_5(\tau,\tau^p)=\int d\omega |\alpha_j(\omega)||\alpha^{d*}_2(\omega)|e^{[i\tau^p(\omega-\omega_0)-i\tau(\omega-\omega^d_0)]}.
\label{eq:J2}
\end{eqnarray}

The results are given in Figs. 2 and 3, where we plot $g^{(2)}$ for $\omega^d_0=4.5\times 10^{14}$ and $\omega_0=5\times 10^{14}$ (namely a detector-photon frequency mismatch of $0.5\times 10^{14}$ or 10\%), and where we plot for five different photon source pair separation: $\tau^p=0,1.5,2.5, 5$ and $10$ $[10^{-12}\,s]$.

\begin{figure}[b]%
\includegraphics[width=\columnwidth]{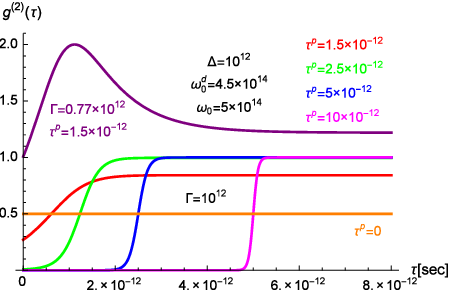}%
\caption{
$g^{(2)}(\tau)$. For $\tau^p=0$, one always finds $g^{(2)}(\tau)=1/2$. For large $\tau^p$ we observe behavior characteristic of single photon sources, $g^{(2)}(\tau=0)=0$. We also observe that different from sources such a chaotic, coherent or single photon emitter, $g^{(2)}(\tau\rightarrow \infty)\ne 1$.
For $\Gamma=0.77\times 10^{12}$, we observe $g^{(2)}>1$, namely bunching. Changing to $\Gamma=1\times 10^{12}$ bunching disappears exhibiting the fact that it is sensitive to detector bandwidth. For $\tau^p=10^{-12}(10^{-13})$ we observe $g^{(2)}$ in the order of $10(10^4)$, depending on $\Gamma$. Similarly we get extremely high values of $g^{(2)}$ as the detector-field frequency mismatch grows. This is so as extremely small detection probabilities create a negligible denominator in Eq. (\ref{eq:new7}), enabling the very large $g^{(2)}$ values. This is qualitatively similar to the experimental observations in Ref. \cite{OneOverN}.
}%
\label{pg22a}%
\end{figure}

\begin{figure}[b]%
\includegraphics[width=\columnwidth]{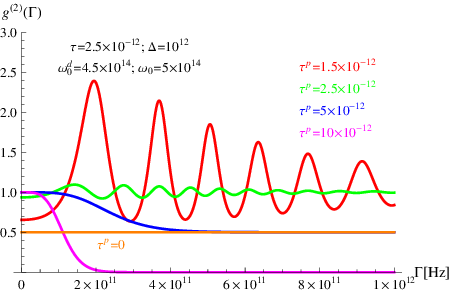}%
\caption{
$g^{(2)}(\Gamma)$. We observe oscillations in $g^{(2)}(\Gamma)$, where these oscillations with varying periods in $\Gamma$, persist for $0<\omega^d_0<\omega_0$ (the oscillation frequency increases with the frequency mismatch $\omega^d_0-\omega_0$). For $\omega^d_0=4.875\times 10^{14}$ one oscillation may be observed, determining the smallest mismatch required for the observation of this phenomenon. The oscillations may be observed up to $\Gamma=5\times 10^{12}$ so that the Heisenberg temporal uncertainty of the detector is smaller than all other time scales in the system. See text for additional features.
}%
\label{pg22c}%
\end{figure}


Let us emphasize some of the observed features. As expected, for $\tau^p=0$, $g^{(2)}(\tau=0)=1/2$. Furthermore, in Fig. \ref{pg22a} we see that well separated states (namely, $\tau^p>1/\Delta$ so that the two photons are prepared with no overlap) give the same statistics as single atom emitters [i.e. $g^{(2)}(\tau\rightarrow 0)=0$]. This is also to be expected. Also not surprising is the fact that in Fig. \ref{pg22c} we see that for $\tau^p=5\times 10^{-12}$ very slow detectors (small $\Gamma$) are not able to turn well separated pairs into indistinguishable pairs thereby erasing their orthogonal preparation. Finally, as expected theoretically for the case of squeezed vacuum \cite{Loudon}, and as observed in experiment \cite{Emmanuel2009,Emmanuel2012,OneOverN}, we find bunching (namely $g^{(2)}>1$ in Fig. \ref{pg22a}). The range of values we find is extraordinarily high and may reach $g^{(2)}=10-10^4$ for $\tau^p=10^{-12}-10^{-13}$. Similarly we get extremely high values of $g^{(2)}$ as the detector-field frequency mismatch grows. This is so as extremely small detection probabilities create a negligible denominator in Eq. (\ref{eq:new7}), enabling the very large $g^{(2)}$ values. This is qualitatively similar to the experimental observations in Ref. \cite{OneOverN}.

More surprising is the fact that from both Fig. \ref{pg22a} and Fig. \ref{pg22c} one finds that perfectly overlapped (indistinguishable) states ($\tau^p=0$) give $g^{(2)}(\tau)=1/2$ for any set of parameters, unlike the traditional sources which are typically discussed (thermal, coherent, and single atom) which all have $g^{(2)}(\tau\rightarrow \infty)=1$, and unlike what is observed in Fig. \ref{PhotonCounting}. Also different from typical sources is the fact that when $g^{(2)}(\tau)$ tends towards an asymptotic value, it is not always $1$; one can clearly observe in Fig. \ref{pg22a} that when the photons are partially overlapping, $g^{(2)}(\tau\rightarrow \infty)\ne 1$. This is again contrary to what has been observed in Fig. \ref{PhotonCounting}.

Finally, it seems clear from Fig. \ref{pg22c} that by changing the bandwidth of the detector we are able to change the value of  $g^{(2)}$. One may say that we should not be surprised, since for example, only fast electronics can see the $g^{(2)}=2$ of a thermal source (otherwise its $g^{(2)}=1$). But in this case the explanation is simple: in a thermal source there are strong intensity fluctuations which cause the emitted state to be highly bunched for short times (e.g. \cite{Loudon}). Hence, only if you have fast electronics can you see this bunching.  Similarly, if you have a single atom source which you continuously excite, only if you have fast electronics can you notice that at short times $g^{(2)}=0$. What would be the explanation in the case of a 2-photon source? We would expect that a higher temporal resolution (larger $\Gamma$) would lead for separated photons ($\tau^p\ne0$) to $g^{(2)}=0$, but in some cases it only leads to $g^{(2)}=1/2$ (see $\tau^p=5\times 10^{-12}$, where the same behavior is also seen for $\tau=5\times 10^{-12}$ and $\tau^p=10\times 10^{-12}$).
This is counter intuitive as $g^{(2)}=1/2$ is, as Eq. (\ref{eq:g2b}) shows us, the hall mark of indistinguishability, and if anything, higher temporal resolution should lead to distinguishability. Even more puzzling, we find an oscillatory dependence of $g^{(2)}$ on the bandwidth $\Gamma$. Oscillations have been previously observed as a function of time \cite{Rempe} but not as a function of $\Gamma$. While there may be some connection between the two, it is not straightforward, and thus a simple physical origin of the oscillations observed here remains outside our reach at present.

Let us now go a step further and use a more complex form of the operators \cite{YoniYair}. In the above calculations, the photon spectral and temporal amplitudes had a Gaussian form and thus had an infinite extent in time. A similar distribution was used for the detector response. One may claim that for a physical picture, one should ensure causality \cite{causal}. We now move to reproduce our calculations with all the operators having the causal form. In the following, we shall use the form appearing in Ref. \cite{Elena1} (see further derivation in the Appendices).

We shall now re-do the relevant equations. Let us first plot Fig. 1a. Eqs. (\ref{eq:incoming1}) and (\ref{eq:alpha2}) now take the form
\begin{eqnarray}
\hat{A}_{1i}^\dag=\int d\omega \alpha_i^*(\omega)\hat{a}_1^\dag(\omega)~~~~
\hat{A}_{1j}^\dag=\int d\omega \alpha_j^*(\omega)\hat{a}_1^\dag(\omega),
\label{eq:incoming1a}
\end{eqnarray}
where we take the spectral amplitude of both photons to be (appendix I)
\begin{eqnarray}
\alpha_{i,j}(\omega)=i\sqrt{\frac{\Delta}{2\pi}}\text{exp}[-i\omega t_{i,j}]\times
\frac{1}{\omega-\omega_0+i\Delta/2},
\label{eq:alphaa}
\end{eqnarray}
and where $\omega_0$ is, as before, the photon center frequency, and $\Delta$ is its frequency spectrum width.

The commutation relation of Eq. (\ref{eq:incoming7b}) now becomes
\begin{eqnarray}
[\hat{A}_{1i},\hat{A}_{1j}^\dag]=\nonumber\\
\frac{\Delta}{2\pi}\int d\omega \int d\omega' \frac{i}{\omega-\omega_0+i\Delta/2} \left[\frac{i}{\omega'-\omega_0+i\Delta/2}\right]^*\times \nonumber\\
\left[\hat{a}_1(\omega),\hat{a}^\dag_1(\omega')\right]e^{-i\omega t_i} e^{i\omega' t_j}=\nonumber\\
\int d\omega  \frac{1}{(\omega-\omega_0)^2+\Delta^2/4} e^{i\omega \tau}=\nonumber\\
e^{-i\omega_0\tau}e^{-|\Delta\tau|/2}=J(\tau),
\label{eq:incoming7ba}
\end{eqnarray}
where $\tau=t_j-t_i$.

We now have, as before, $[\hat{A}_{1i},\hat{A}_{1i}^\dag]=[\hat{A}_{1j},\hat{A}_{1j}^\dag]=1$ and $[\hat{A}_{1i},\hat{A}_{1j}^\dag]=[\hat{A}_{1j},\hat{A}_{1i}^\dag]^*=J$, and using
the second order correlation of Eq. (\ref{eq:dis3}) we plot the result in Fig. 4.

\begin{figure}[b]%
\includegraphics[width=\columnwidth]{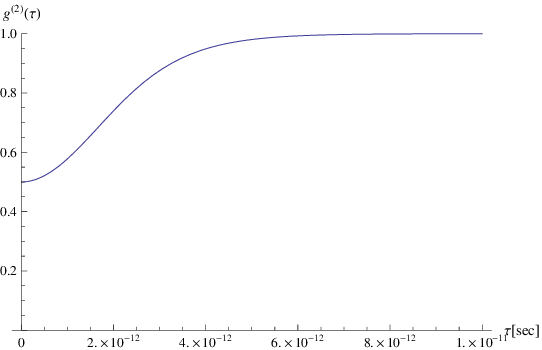}%
\caption{
$g^{(2)}(\tau)$ according to Eq. (\ref{eq:dis3}) with $J(\tau)$ taken from Eq. (\ref{eq:incoming7ba}). Fig. 1 is qualitatively reproduced.
}%
\label{PhotonCountinga}%
\end{figure}

\begin{figure}[b]%
\includegraphics[width=\columnwidth]{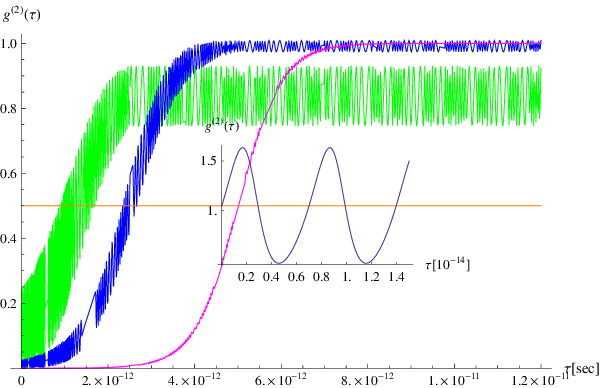}%
\caption{
$g^{(2)}(\tau)$. The color codes are as before, and so are the values of other parameters. Different from the previous model, here oscillations may be observed. It is not clear if their origin is numeric. Inset: the $\tau^p=1.5\times 10^{-12}$ plot has a large oscillation amplitude and was thus separated. Here $\tau=0$ means $2\times 10^{-12}$ seconds.
}%
\label{fig5}%
\end{figure}

\begin{figure}[b]%
\includegraphics[width=\columnwidth]{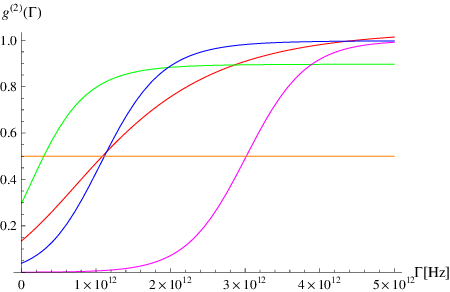}%
\caption{
$g^{(2)}(\Gamma)$. The color codes are as before, and so are the values of other parameters.
}%
\label{fig6}%
\end{figure}

\begin{figure}[b]%
\includegraphics[width=\columnwidth]{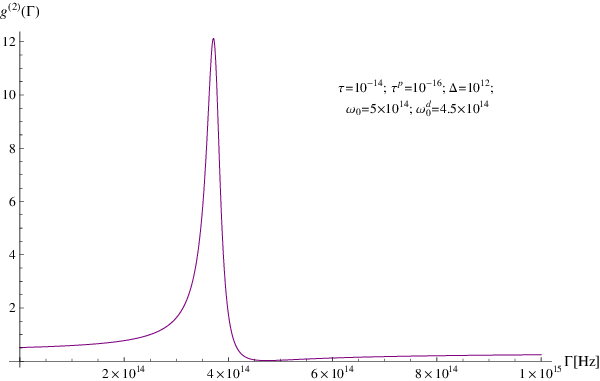}%
\caption{
$g^{(2)}(\Gamma)$. We see that for some parameters we again find a non-monotonic behavior. As in Fig. 3, the non-monotonic behavior is observed when $g^{(2)}_{max}>1$.
}%
\label{fig7}%
\end{figure}

In Figs. \ref{fig5}-\ref{fig7} we now redo Figs. 2 and 3. Based on the new J values (appendix II) we rewrite Eq. (\ref{eq:J2}) and recalculate Eq. (\ref{final}). We again find that bunching occurs in some region of the parameter space, and similarly we again find that the second order coherence function may oscillate as a function of the detector bandwidth.

Finally, let us attempt to interpret the observed oscillations in the realm of complementarity. Complementarity, meaning that ``coherence" is dependent on measurement \cite{Walther}, typically presents itself through a monotonic relation $V^2+D^2\le 1$ (where $V$ is the visibility of the fringes, or $g^{(1)}$ value, and $D$ the distinguishability), in which, as distinguishability goes up, the value of $g^{(1)}$ goes down \cite{Greenberger,Englert}.
One has to be cautious in utilizing the language of complementarity, but if one takes the temporal resolution of our detector as the distinguishability offered by our ``which path" measurement, as at high resolution it can tell which particle came first, we find that in contrast to the $g^{(1)}$ complementarity, the $g^{(2)}$ complementarity may oscillate.

Most of the experimental work done on this issue followed the original "Welcher Weg" (WW) notion of the effect of making a measurement in the spatial (position) basis. These interferometer (first-order coherence, $g^{(1)}$) single particle experiments (e.g. \cite{Mandel0,heiblum1,kwiat,heiblum2}) clearly showed that making a WW measurement on a spatial superposition (wave-like) state, ``collapses" the state into a localized (particle-like) state. In fact, the WW measurement does not have to be done prior to the final measurement of an interference pattern. In a double slit experiment, if the detector on which the interference pattern is to be recorded, has sufficient angular resolution so that it can tell from which direction or slit the particle came, no interference pattern will be recorded.

By utilizing a two particle state and measuring its intensity correlation we in fact have made an analogy to the above experiment. Similar to the latter $g^{(1)}$ experiment, our present scheme requires one detector which will acquire both the final value of the coherence function and the additional WW knowledge (in our case, which particle arrived first), where we control the level of acquired additional knowledge by changing the detector bandwidth.

Recent works suggest that complementarity may perhaps present itself in more complex forms than previously thought \cite{Menzel1,Menzel2,Yoon}.
In a broad sense, one may say that complementarity is the effect of the measurement of one observable on the value of some other observable. In this sense, we have shown that this broad complementarity may oscillate.

This work focused on the theoretical description of a new effect, and as such the experimental feasibility is beyond the scope of this paper and will be presented elsewhere. Nevertheless, we note two important points: first, the results remain the same if all frequencies are similarly scaled down so that the experiment may be done, for example, also in the micro-wave regime \cite{Lang}; second, for a minimal field-detector mismatch enabling the observation of a single oscillation (Fig. \ref{pg22c}), the probability of a detection, which is nothing else but the expectation values appearing in the denominator of Eq. (\ref{eq:new7}), is a few percent. Hence, the experiment seems to be within reach.

To conclude, we have analyzed an $n=2$ state, in the situation of a field-detector frequency mismatch. We observe anomalous features of the $g^{(2)}$ function and expect the experimental and theoretical investigation of these features to yield new insight. In addition, we believe that making an analogy between this experiment and WW experiments, opens up new possibilities in exploring complementarity. Let us note that we have used in this work simplified models, which may and should be further improved if one is to fully understand the region of validity and the implications of the results.

I gratefully acknowledge discussions with Claude Cohen-Tannoudji, Elena Del Valle, Carsten Henkel, Andrej Singer and Ulf Lorenz. I am also thankful to Yair Margalit, Daniel Rohrlich, Judy Kupferman and Yonathan Japha of the atom chip group. This work is partly supported by the Miller Institute for Basic Research in Science, University of California Berkeley. \\

APPENDIX I\\

Elena del Valle et. al. \cite{Elena1} introduce the autocorrelation function
$S^{\Gamma}_{\omega}(t)=\langle A_{\omega}^{\dag}(t)A_{\omega}(t)\rangle$ of detection in both
time and frequency, where $S^{\Gamma}_{\omega}(t)$ is interpreted as a spectral density for detection
at a given time and for a given bandwidth of the detector ($\Gamma$). This is our starting point.
The operator $A_{\omega}(t)$ is given in terms of the field operator $a(t)$ by a causal form
\begin{eqnarray}
A_{\omega_0}(t)=\sqrt{\frac{\Gamma}{2\pi}}\int_{-\infty}^t dt' e^{-i\omega_0 (t-t')}e^{-\Gamma (t-t')/2}a(t')
\label{eq:new2}
\end{eqnarray}
The field $a(t')$ can be expressed in terms of its Fourier components as
\begin{eqnarray}
a(t')=\int_{-\infty}^{\infty}d\omega a(\omega)e^{-i\omega t'}.
\label{eq:new2a}
\end{eqnarray}
By substituting this in the definition of $A_{\omega_0}(t)$ we obtain
\begin{eqnarray}
A_{\omega_0}(t)=\sqrt{\frac{\Gamma}{2\pi}}\int_{-\infty}^{\infty}d\omega e^{-i\omega t}\int_0^{\infty} d\tau e^{i(\omega-\omega_0+i\Gamma/2) \tau}a(\omega)=\nonumber\\
i\sqrt{\frac{\Gamma}{2\pi}}\int_{-\infty}^{\infty}d\omega e^{-i\omega t}\frac{1}{\omega-\omega_0+i\Gamma/2}a(\omega) \label{eq:new2b}
\end{eqnarray}\\

APPENDIX II\\

In the following we specifically calculate J1-J5, following the derivation presented in appendix III.

\begin{eqnarray}
J_1(\tau^p)=[\hat{A}_{1i},\hat{A}_{1j}^\dag]=\nonumber\\
\frac{\Delta}{2\pi}\int d\omega \int d\omega' \frac{i}{\omega-\omega_0+i\Delta/2} \left[\frac{i}{\omega'-\omega_0+i\Delta/2}\right]^*\times \nonumber\\
\left[\hat{a}_1(\omega),\hat{a}^\dag_1(\omega')\right]e^{-i\omega t_i} e^{i\omega' t_j}=\nonumber\\
\int d\omega  \frac{1}{(\omega-\omega_0)^2+\Delta^2/4} e^{i\omega \tau^p}=\nonumber\\
e^{-i\omega_0\tau^p}e^{-|\Delta\tau^p|/2}
\label{eq:newJ1}
\end{eqnarray}

\begin{eqnarray}
J_2=[\hat{A}_{1i},\hat{D}_{11}^\dag]=\nonumber\\
\frac{\sqrt{\Delta\Gamma}}{2\pi}\int d\omega \int d\omega' \frac{i}{\omega-\omega_0+i\Delta/2} \left[\frac{i}{\omega'-\omega^d_0+i\Gamma/2}\right]^*\times \nonumber\\
\left[\hat{a}_1(\omega),\hat{a}^\dag_1(\omega')\right]e^{-i\omega t_i} e^{i\omega' t_1}=\nonumber\\
\frac{\sqrt{\Delta\Gamma}}{2\pi}\int d\omega  \frac{1}{\omega-\omega_0+i\Delta/2} \left[\frac{1}{\omega-\omega^d_0+i\Gamma/2}\right]^* =\nonumber\\
C\times [2\theta(0)]
\label{eq:newJ2}
\end{eqnarray}

\begin{eqnarray}
J_3(\tau)=[\hat{A}_{1i},\hat{D}_{12}^\dag]=\nonumber\\
\frac{\sqrt{\Delta\Gamma}}{2\pi}\int d\omega \int d\omega' \frac{i}{\omega-\omega_0+i\Delta/2} \left[\frac{i}{\omega'-\omega^d_0+i\Gamma/2}\right]^*\times \nonumber\\
\left[\hat{a}_1(\omega),\hat{a}^\dag_1(\omega')\right]e^{-i\omega t_i} e^{i\omega' t_2}=\nonumber\\
\frac{\sqrt{\Delta\Gamma}}{2\pi}\int d\omega  \frac{1}{\omega-\omega_0+i\Delta/2} \left[\frac{1}{\omega-\omega^d_0+i\Gamma/2}\right]^* e^{i\omega \tau}=\nonumber\\
C\times [e^{i\omega_0 \tau}\theta(-\tau)e^{\Delta \tau/2}+\nonumber\\
e^{i\omega^d_0\tau}\theta(\tau)e^{-\Gamma \tau/2}]
\label{eq:newJ3}
\end{eqnarray}

\begin{eqnarray}
J_4(\tau^p)=[\hat{A}_{1j},\hat{D}_{11}^\dag]=\nonumber\\
\frac{\sqrt{\Delta\Gamma}}{2\pi}\int d\omega \int d\omega' \frac{i}{\omega-\omega_0+i\Delta/2} \left[\frac{i}{\omega'-\omega^d_0+i\Gamma/2}\right]^*\times \nonumber\\
\left[\hat{a}_1(\omega),\hat{a}^\dag_1(\omega')\right]e^{-i\omega t_j} e^{i\omega' t_1}=\nonumber\\
\frac{\sqrt{\Delta\Gamma}}{2\pi}\int d\omega  \frac{1}{\omega-\omega_0+i\Delta/2} \left[\frac{1}{\omega-\omega^d_0+i\Gamma/2}\right]^* e^{-i\omega \tau^p}=\nonumber\\
C\times [e^{-i\omega_0 \tau^p}\theta(\tau^p)e^{-\Delta \tau^p/2}+\nonumber\\
e^{-i\omega^d_0\tau^p}\theta(-\tau^p)e^{\Gamma \tau^p/2}]
\label{eq:newJ4}
\end{eqnarray}

\begin{eqnarray}
J_5(\tau,\tau^p)=[\hat{A}_{1j},\hat{D}_{12}^\dag]=\nonumber\\
\frac{\sqrt{\Delta\Gamma}}{2\pi}\int d\omega \int d\omega' \frac{i}{\omega-\omega_0+i\Delta/2} \left[\frac{i}{\omega'-\omega^d_0+i\Gamma/2}\right]^*\times \nonumber\\
\left[\hat{a}_1(\omega),\hat{a}^\dag_1(\omega')\right]e^{-i\omega t_j} e^{i\omega' t_2}=\nonumber\\
\frac{\sqrt{\Delta\Gamma}}{2\pi}\int d\omega  \frac{1}{\omega-\omega_0+i\Delta/2} \left[\frac{1}{\omega-\omega^d_0+i\Gamma/2}\right]^* e^{i\omega (\tau-\tau^p)}=\nonumber\\
C\times [e^{-i\omega_0 (\tau^p-\tau)}\theta(\tau^p-\tau)e^{-\Delta (\tau^p-\tau)/2}+\nonumber\\
e^{-i\omega^d_0(\tau^p-\tau)}\theta(\tau-\tau^p)e^{\Gamma (\tau^p-\tau)/2}]
\label{eq:newJ5}
\end{eqnarray}
where $C=\frac{\sqrt{\Delta\Gamma}}{(\Delta+\Gamma)/2+i(\omega^d_0-\omega_0)}$.\\

APPENDIX III\\

We start from the simple integral
\begin{eqnarray}
\int_0^{\infty} dt e^{i\omega t}e^{-\Delta t/2}=\frac{i}{\omega+i\Delta/2}
\label{eq:appendix3}
\end{eqnarray}

This implies that the functions $i/(\omega+i\Delta/2)$ and $\theta(t)e^{-\Delta t/2}$ are a Fourier transform pair,
such that
\begin{eqnarray}
I(\Delta,t)=\int_{-\infty}^{\infty} \frac{d\omega e^{-i\omega t}}{\omega+i\Delta/2}=-2\pi i\theta(t)e^{-\Delta t/2}\nonumber\\
I(-\Delta,t)=\int_{-\infty}^{\infty} \frac{d\omega e^{-i\omega t}}{\omega-i\Delta/2}=2\pi i\theta(-t)e^{\Delta t/2}
\label{eq:appendix3a}
\end{eqnarray}
where $\theta(t)=1$ for $t>0$ and zero elsewhere.

Consider the following integral
\begin{eqnarray}
\int_{-\infty}^{\infty} \frac{d\omega e^{-i\omega t}}{\omega-\omega_1+i\Delta_1/2}\frac{1}{\omega-\omega_2-i\Delta_2/2}=\nonumber\\
\frac{1}{(\omega_2-\omega_1)-i(\Delta_1+\Delta_2)/2}\times\nonumber\\
\left[e^{-i\omega_1 t}I(\Delta_1,t)-e^{-i\omega_2 t}I(-\Delta_2,t)\right]=\nonumber\\
\frac{2\pi}{(\Delta_1+\Delta_2)/2+i(\omega_2-\omega_1)}\times\nonumber\\
\left[e^{-i\omega_1 t}\theta(t)e^{-\Delta_1 t/2}+\theta(-t)e^{\Delta_2 t/2}\right]
\label{eq:appendix3b}
\end{eqnarray}

The integral identity that follows is
\begin{eqnarray}
J(\omega_1,\omega_2,\Delta_1,\Delta_2,\tau)=\nonumber\\
\frac{\sqrt{\Delta_1\Delta_2}}{2\pi}\int_{-\infty}^{\infty}
\frac{d\omega e^{-i\omega\tau}}{(\omega-\omega_1+i\Delta_1/2)(\omega-\omega_2-i\Delta_2/2)}=\nonumber\\
\frac{\sqrt{\Delta_1\Delta_2}}{(\Delta_1+\Delta_2)/2+i(\omega_2-\omega_1)}\times\nonumber\\
\left[e^{-i\omega_1 \tau}\theta(\tau)e^{-\Delta_1 \tau/2}+
e^{-i\omega_2\tau}\theta(-\tau)e^{\Delta_2 \tau/2}\right]
\label{eq:appendix3c}
\end{eqnarray}

When $\omega_1=\omega_2=\omega_0$ and $\Delta_1=\Delta_2=\Delta$ we obtain
$J=e^{-i\omega_0 \tau}e^{-\Delta|\tau|/2}$ as in Eq. (\ref{eq:incoming7ba}).

\bibliographystyle{apsrev4-1} 

\begin{thebibliography}{10}%




\bibitem{HBT1954} R. Hanbury Brown and R. Q. Twiss, Phil. Mag. {\bf 45}, 663 (1954).

\bibitem{Fano1961} U. Fano, Am. J. Phys. {\bf 29}, 539 (1961).

\bibitem{Baym1969} G. Baym, {\it Lectures on quantum mechanics}, Benjamin Pub., New-York (1969).

\bibitem{HBT1974} R. Hanbury Brown and R. Q. Twiss, {\it The intensity interferometer}, Taylor and Frances, London (1974).

\bibitem{Hong1987} C. K. Hong, Z. Y. Ou and L. Mandel, Phys. Rev. Lett. {\bf 59}, 2044 (1987).

\bibitem{Ou1989} Z. Y. Ou and L. Mandel, Phys. Rev. Lett. {\bf 62}, 2941 (1989).

\bibitem{Pittman1996} T. B. Pittman et al., Phys. Rev. Lett. {\bf 77}, 1917 (1996).

\bibitem{Mandel} L. Mandel, Review of Modern Physics {\bf 71}, S274 (1999).

\bibitem{Boyd2008} Anand Kumar Jha, Malcolm N. O'Sullivan, Kam Wai Clifford Chan, and Robert W. Boyd, Temporal coherence and indistinguishability in two-photon interference effects, Phys. Rev. A {\bf 77}, 021801(R) (2008).

\bibitem{Scully} Marlan O. Scully and M. Suhail Zubairy, {\it Quantum optics}, Cambridge  university press (1996).

\bibitem{ron} R. Folman, Two-particle quantum transmission, arXiv:1201.3111 (2012); (Invited paper) Proceedings of SPIE, Volume 8518, 8518 0H (2012).

\bibitem{shih} Morton H. Rubin, David N. Klyshko, Y. H. Shih and A. V. Sergienko, Theory of two-photon entanglement in type-II optical parametric
down-conversion, Phys. Rev. A {\bf 50}, 5122 (1994).

\bibitem{Rempe} T. Legero et al., Quantum Beat of Two Single Photons, Phys. Rev. Lett. {\bf 93}, 070503 (2004).

\bibitem{Ou2006} Z. Y. Ou, Temporal distinguishability of an N-photon state and its characterization by quantum interference, Phys. Rev. A {\bf 74}, 063808 (2006).

\bibitem{Emmanuel2009} F. Boitier, A. Godard, E. Rosencher and C. Fabre, Measuring photon bunching at ultrashort timescale by two photon absorption in semiconductors, Nature Physics {\bf 5}, 267 (2009).

\bibitem{Emmanuel2012} Fabien Boitier, Antoine Godard, Nicolas Dubreuil, Philippe Delaye, Claude Fabre, and Emmanuel Rosencher, Two-Photon Counting Interferometry, Phys. Rev. A {\bf 87}, 013844 (2013).

\bibitem{Elena1} Elena d. Valle et al., Theory of Frequency-Filtered and Time-Resolved N-Photon Correlations, Phys. Rev. Lett. {\bf 109}, 183601 (2012).

\bibitem{Elena2} Elena d. Valle, Distilling one, two and entangled pairs of photons
from a quantum dot with cavity QED effects and
spectral filtering, New J. of Phys. {\bf 15}, 025019 (2013).

\bibitem{Elena3} Alejandro Gonzalez-Tudela et al., Two-photon spectra of quantum emitters, arXiv:1211.5592v1 (2012).

\bibitem{scully} J. Brendel et al., A beam splitting experiment with correlated photons, Europhys. Lett. {\bf 5}, 223 (1988).

\bibitem{grainger} J. Beugnon et al., Quantum interference between two single photons
emitted by independently trapped atoms, Nature {\bf 440}, 779 (2006).

\bibitem{BS} R. Loudon, Phys. Rev. A {\bf 58}, 4904 (1998).

\bibitem{Loudon} R. Loudon, {\it The quantum theory of light}, Third edition, Oxford university press.

\bibitem{OneOverN} Dmitry A. Kalashnikov, Si Hui Tan, Maria V. Chekhova, and Leonid A. Krivitsky, Optics Express {\bf 19} 9352 (2011).

\bibitem{generality} An addition of some time delay $t_0$ between the production of the first photon and the detection of the first photon, would introduce a term $exp[i\omega t_0]$. For all relevant values of $\omega$ in our system, one can find a short enough $t_0$, e.g. femto-second, such that this term is practically a constant.

\bibitem{YoniYair} The analytical form of the complex J functions was developed by Yonathan Japha and the numerics were made by Yair Margalit.

\bibitem{causal} Malin Premartne and Govind P. Agrawal, {\it Light propagation in gain media}, Sec. 1.4, Cambridge university press (2011).

\bibitem{Walther} Marlan O. Scully, Berthold-Georg Englert and Herbert Walther, Quantum optical tests of complementarity, Nature {\bf 351}, 111 (1991).

\bibitem{Greenberger} D. M. Greenberger, A. Yasin, Phys. Lett. A. {\bf 128}, 391394 (1988).

\bibitem{Englert} B.-G. Englert, Phys. Rev. Lett. {\bf 77}, 2154 (1996).

\bibitem{Menzel1} R. Menzel, D. Puhlmann, A. Heuer, W.P. Schleich, Proc. Natl. Acad. Sci. USA {\bf 109}, 9314 (2012).

\bibitem{Menzel2} R. Menzel, A. Heuer, D. Puhlmann, K. Dechoum, M. Hillery, M.J.A. Sp\"ahne, W.P. Schleich, J. Mod. Optics (2012), DOI:10.1080/09500340.2012.746400.

\bibitem{Yoon} Young-Sik Raa, Malte C. Tichyb, Hyang-Tag Lima, Osung Kwona, Florian Mintertb, Andreas Buchleitnerb, and Yoon-Ho Kima, PNAS {\bf 110} (4), 1227 (2012).

\bibitem{Mandel0} X. Y. Zou, L. J. Wang and L. Mandel, Induced coherence and indistinguishability in optical
interference. Phys. Rev. Lett. {\bf 67}, 318 (1991).

\bibitem{heiblum1} E. Buks, R. Schuster, M. Heiblum, D. Mahalu, V. Umansky, Dephasing in electron interference by a 'which-path' detector, Nature {\bf 391}, 871 (1998).

\bibitem{kwiat} Peter D. D. Schwindt, Paul G. Kwiat, and Berthold-Georg Englert, Quantitative wave-particle duality and nonerasing quantum erasure, Phys. Rev. A {\bf 60}, 4285 (1999).

\bibitem{heiblum2} D. Rohrlich, Mesoscopic interferometers for electron waves, Optics and Spectroscopy {\bf 99}, 503 (2005).

\bibitem{Lang} C. Lang et al., Probing Correlations, Indistinguishability and Entanglement in Microwave Two-Photon Interference, arXiv:1301.4458v1 (2013).

\end{thebibliography}

%

\end{document}